\documentclass[fleqn,twoside]{article}
\usepackage{espcrc2}

\usepackage{graphicx}
\newcommand{\AmS}{{\protect\the\textfont2
  A\kern-.1667em\lower.5ex\hbox{M}\kern-.125emS}}

\title{Final state interactions in electron scattering
       at high missing energies and momenta}
\author{C. Barbieri\address{TRIUMF, 4004 Wesbrook Mall, Vancouver, 
                                British Columbia, Canada V6T 2A3}%
                   \address{Gesellschaft f\"ur Schwerionenforshung,
                                Planckstr. 1, 64291, Darmstadt, Germany}%
         \thanks{Email: {\tt C.Barbieri@gsi.de}}%
         \thanks{Present address: Gesellschaft f\"ur Schwerionenforshung,
                                Planckstr. 1, 64291, Darmstadt, Germany}
        }
       
\begin{document}

\begin{abstract}

Calculations of two-step rescattering and pion emission for the ${}^{12}{\rm C}(e,e'p)$
reaction at very large missing energies and momenta are compared with recent data from
TJNAF.
 For parallel kinematics, final state interactions are strongly
reduced by kinematical constraints. A good agreement between calculation
and experiment is found for this kinematics when one admits the presence
of high momentum components in the nuclear wave function. \\

%\vspace{.5cm}

\end{abstract}

\maketitle

In recent years, electron scattering experiments have been possible with
kinematics that involve large energies and momentum transfered, as for
example in studies of nuclear transparency.  In this regime, the final state
interactions (FSI) for the knock out of a nucleon are usually treated
employing Glauber inspired
calculations~\cite{Ciofi05,Strikman05,Schiavilla05,BenharOka}.
 Such techniques, together with the experience gained in related nuclear
structure studies, provide a useful starting point to pursue accurate
predictions of neutrino-nucleus interactions at high energy. This was pointed
out by various contributions to this conference.

At scattering energies of the order of GeV, the leptoninc probe
can resolve the high-momentum tail of the spectral function generated
by short-range (central and tensor) correlations (SRC).
This represent about 10-20\% of the total spectral
strength~\cite{PandSick97,DB04} and is found along a ridge in
the momentum-energy plane ($k$-$E$) which spans a region of  
several hundred MeV/c (MeV)~\cite{Ciofi90,WHA,Benhar}.
 This corresponds to large missing momenta ($p_m$) and energies ($E_m$)
in knock out cross sections.
This contribution to the spectral function is also responsible
for most of the binding energy of nuclear systems~\cite{Wim03}.
The main characteristics predicted by these calculations are confirmed by
recent experimental data~\cite{danielaPRL,Frick04,RoheOka},
which will be considered further below.

 Locating this strength experimentally, at both large $E_m$ and $p_m$, is
difficult because it is spread over an energy range of several hundred MeV,
so the total density of the spectral function is very low. 
 In this energy regime multi-nucleon processes, beyond the direct knock out,
are possible~\cite{Takaki89} and can induce large shifts in the
missing energies and momenta, moving strength to regions where the direct signal
is much smaller and therefore submerging it.
 As it will be discussed below, the effects of FSI become larger and less
controllable when the transverse structure functions that enter
the expression of the ($e,e'p$) cross section dominate the longitudinal one.
This trend is predicted by several theoretical studies~\cite{Takaki89,VivianPV,%
NikolaevGL,NikolaevNoSRC_and_Ciofi4He,RyckebuschO16}.
%%
%Interference effects between FSI and initial state correlations (IC)
% can also play a role~\cite{NikolaevNoSRC_and_Ciofi4He}.
%%
% The results of Refs.~\cite{VivianPV,Ryckebusch03} suggests that multiple
%rescattering contributions (more than two-steps) are relatively small
%in light nuclei and when parallel kinematics are considered%
%
The issue of how to control FSI in lepton scattering experiments has been
discussed in detail in Ref.~\cite{IngoElba,E97proposal} for the particular case
of kinematics sensitive to the SRC tail of the nuclear spectral function.
 A Monte Carlo simulation and kinematical arguments led
to the suggestion that the best chance for an identification of SRC
occurs in parallel kinematics
\footnote{In this work we refer to `parallel' and `perpendicular' kinematics
in terms of the angle between the momentum transfered by the leptonic current
${\bf q}$ and the momentum of the initial nucleon ${\bf p}_i=-{\bf p}_m$
(as opposed to the detected proton ${\bf p}_f$).
 This definition is more restrictive in the limit of high momentum transfer,
where ${\bf q}$ and ${\bf p}_f$ tend to be collinear anyhow.}.
The latter also tend to be less sensitive to meson exchange currents
(MEC) -- which involve transverse excitations -- and are cleaner due
to the high momentum that is required for the detected proton.
 New data were subsequently taken by the \hbox{E97-006} collaboration
at Jefferson Lab~\cite{danielaDiss,danielaPRL,Frick04} for
a set of nuclei ranging from carbon to gold. 
Both optimal (parallel) and perpendicular kinematics were used,
to provide useful data for investigating the reaction mechanism.

At the energy regime of the JLab experiment,
the main contribution to FSI, below the meson production threshold, is
identified with two-step rescattering. This has been studied recently
in Refs.~\cite{BaLap04,BaRSL05} using a semiclassical model, which has
the advantage of describing the distortion due to FSI in terms of the full
one-hole spectral function.
 This accounts for both the momentum and the energy distribution 
of the original correlated strength, which is of importance for the
proper description of the response~\cite{SpectFunctPRL}.
In this letter we extend the analysis of Ref.~\cite{BaRSL05} by including
the contributions of $\pi$-emission and compare with the data
of Ref.~\cite{danielaDiss}.% for the nucleus ${}^{12}{\rm C}$.

% X-sec formulae
%---------------

In the ideal case where only the direct knock out process was relevant,
the $(e,e'p)$ cross section for emitting a proton from a nucleus with
spectral function $S^h_p(k,E)$ would be correctly described by
plane wave impulse approximation (PWIA),
\begin{eqnarray}
  \lefteqn{
     { d^6  \sigma_{PWIA}
     \over
     dE_0 \; d\Omega_{\hat{k}_o}  dE_f \; d \Omega_{\hat{p}_f} }  ~=~
      } \hspace{.1cm}    &~&
\nonumber \\
   &~&    |{\bf p}_f| E_a \; 
          S^h_a(|{\bf q}-{\bf p}_f|,\omega-E_f) \; \sigma^{cc}_{ea}
          \;  {\cal T}  \; \; ,
\label{eq:PWIA}
\end{eqnarray}
where $(E_o,{\bf k}_o)$,  $(E_f,{\bf p}_f)$ and $(\omega,{\bf q})$
represent the four-momenta of the detected electron, the final proton
and the virtual photon, respectively.
 $\sigma^{cc}_{eN}$ is the electron-nucleon cross section.
 The present calculations employed the $\sigma^{cc}$ prescription
discussed in Ref.~\cite{danielaDiss}, which is obtained by using
the on-shell current also for off-shell protons%
\footnote{Preferably one uses a prescription to extrapolate
the on-shell cross section to the off-shell case while preserving energy 
and current conservation.
 The analysis of  several possible prescriptions carried out
in Ref.~\cite{danielaDiss} found that the best agreement between the data
of different kinematics is obtained by avoiding any of the {\em ad hoc}
modifications of the kinematics at the electromagnetic vertex
as suggested in\ Ref.~\cite{deForest}.}.
 In Eq.~(\ref{eq:PWIA}), the transparency factor ${\cal T}$ accounts for the
loss of flux from the direct channel due to the FSI~\cite{BenharOka}.
 However, one still has to correct for the reappearance of strength
through other reaction channels.

In Ref.~\cite{BaRSL05}, we considered the two-step mechanism where the knock
out reaction $(e,e'a)$ for a nucleon $a$ is followed by a scattering process
from another nucleon in the medium, $N'(a,p)N''$. Eventually, leading to the
emission of the detected proton and an undetected nucleon $N''$.
 Three channels are possible, in which $a$
represents either a proton (with $N'=p$ or $n$) or a neutron (with $N'=p$).
The semi-exclusive cross section for these events was calculated
semiclassically as~\cite{BaLap04}
%according to Ref.~\cite{BaLap04} as
%
\begin{eqnarray}
  \lefteqn{
    { d^6  \sigma_{rescatt.}
     \over
     dE_0 \; d\Omega_{\hat{k}_o}  dE_f \; d \Omega_{\hat{p}_f} } 
    }
 \hspace{.2cm}  &~&
\nonumber  \\
 &=&  \sum_{a , N' = 1 , 2 , 3}
    \int d {\bf r}_1 \int d {\bf r}_2 \int_{0}^{\omega} d T_a
     \; \rho_N({\bf r}_1)
\nonumber  \\
 &\;& \times 
    { |{\bf p}_a|E_a \; S^h_a(|{\bf q}-{\bf p}_a|,\omega-E_a) \;
                                                        \sigma^{cc}_{ea}
    \over
    M \; ( {\bf r}_1 - {\bf r}_2 )^2 } 
\nonumber  \\
 &\;& \times 
    g_{aN'}(|{\bf r}_1 - {\bf r}_2|) \;
 P_T(p_a; {\bf r}_1 , {\bf r}_2 ) \;
\rho_{N'}({\bf r}_2) \; 
\nonumber  \\
 &\;& \times 
    { d^3  \sigma_{a N'}
    \over
    dE_f \; d \Omega_{\hat{p}_f} } \;
P_T(p_f; {\bf r}_2 , \infty) \; .
\label{eq:Resc}
\end{eqnarray}
Eq.~(\ref{eq:Resc}) assumes that the intermediate particle $a$ is
generated in (PWIA)
by the electromagnetic current at a point ${\bf r}_1$
inside the nucleus.
Here, $S^h_a(k,E)/M$ is the spectral function of the hit particle
$a$ normalized to one [i.e., $M=N$($Z$) if $a$ is a neutron (proton)].
The pair distribution functions
$g_{aN'}(|{\bf r}_1 - {\bf r}_2|)$ account for the joint probability of
finding a nucleon N' at  ${\bf r}_2$ after the particle $a$ has been struck
at ${\bf r}_1$~\cite{gNN}.
The kinetic energy $T_a$ of the intermediate nucleon $a$
ranges from 0 to the energy $\omega$ transfered by the electron.
The point nucleon densities $\rho_N({\bf r})$ were derived from experimental 
charge distributions by unfolding the proton size~\cite{density_tables}.
The factor $P_T(p; {\bf r}_1 , {\bf r}_2)$ gives the transmission
probability that the struck particle $a$ propagates, without any
interactions, to a second point ${\bf r}_2$. There, it rescatters from
the nucleon $N'$ with cross section $d^3  \sigma_{a N'}$.

The second contribution to the $(e,e'p)$ cross section considered here is
the production of an undetected pion.
 By following a PWIA approach this can be written as
\begin{eqnarray}
  \lefteqn{
    { d^6  \sigma_{\pi emiss.}
     \over
     dE_0 \; d\Omega_{\hat{k}_o}  dE_f \; d \Omega_{\hat{p}_f} } 
    ~=~ 
    \Gamma_v \;  \sum_{a=p,n} \; {\cal T}   \;
    } &~&
 \nonumber  \\
   &\;& \times 
    \int d {\bf k}_\pi \; 
    { W^2 \; |{\bf p}_f|
      \over
    (\omega-E_f-\omega_\pi) \; \omega_\pi \; k_\gamma } 
   \; S^h_a(|{\bf p}_b|,E_b)
\nonumber  \\
 &\;& \times 
  \left[  
    R^a_T + \varepsilon_L \; R^a_L + \varepsilon \; cos(2\phi) 
    \; R^a_{TT} ~ + \right.
\nonumber  \\
  &\;&  \left. ~ + 
     \; \;  \sqrt{2 \varepsilon_L (1 + \varepsilon ) } \; cos(\phi) \; R^a_{LT}
  \right]  \; \; ,
\label{eq:PiEm}
\end{eqnarray}
where, $(\omega_\pi,{\bf k}_\pi)$
represent the four-momenta of the emitted pion, $W$ is the invariant mass of
the emitted pion-nucleus pair,
$\Gamma_v=(\alpha k_\gamma k_o)/[2\pi k_i Q^2 (1- \varepsilon)]$ is the flux
of virtual photon field~\cite{Amaldi79} and the momentum and energy of the bound
nucleon are ${\bf p}_b={\bf q}-{\bf p}_f-{\bf k}_\pi$ and
$E_b=\omega-T_f-T_{\pi}-T_{rec}$.
 The response functions $R^a_{xy}$ refer to the production of a $\pi^0$~($\pi^-$)
when the struck nucleon $a$ is a proton~(neutron).
 In the spirit of the PWIA approach, we employed response functions for
pion production on a free nucleon, evaluated using the MAID
program~\cite{Tiator,MAID-URL}.These depend on $W$ and the angle $\phi$ between
the pion and the photon in the c.o.m. frame.

%  Results
%----------

\begin{figure}[!t]
%%\vspace{.15in}
  \begin{center}
    \includegraphics[width=\linewidth]
                                  {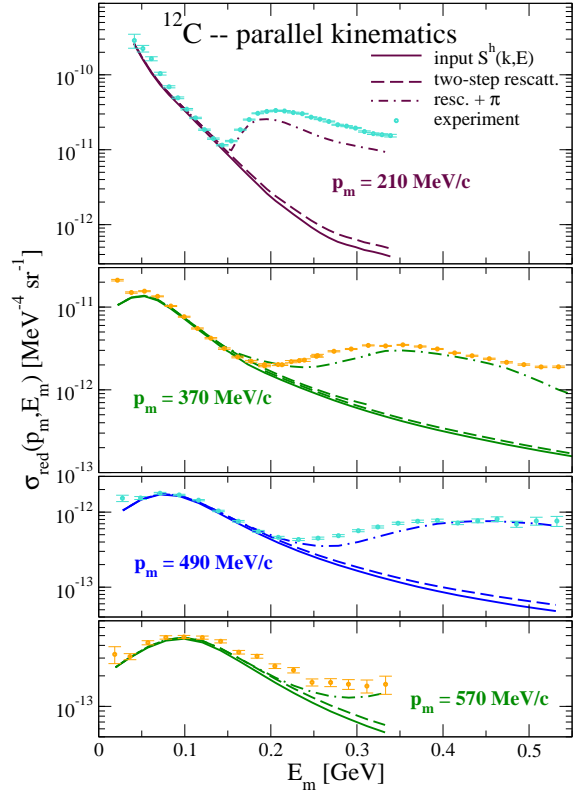}
%\centerline{\psfig{figure=Th_vs_exp_parall.epsi,width=8cm,angle=-90}}
    \caption{ \label{fig:parall_exp}
       (Color online) 
       Theoretical results for the reduced cross section of
       ${}^{12}{\rm C}$  obtained in
      parallel kinematics compared with the experimental
      results of Ref.~\cite{danielaDiss}.
       The full line shows the model spectral function,
      Eq.~(\ref{eq:Shtotal}), employed in the calculations.
       The dashed lines include the additional effects of rescattering,
      Eq.~(\ref{eq:Resc}), while the combined prediction of
      Eq.~(\ref{eq:PWIA}--\ref{eq:PiEm}) is given by the dot-dashed
      lines.
}
\end{center}
\end{figure}

Recently, the experimental strength measured for ${}^{12}{\rm C}$ in
parallel kinematics was compared to theory~\cite{danielaPRL,Frick04}.
For missing energies up to 200~MeV, the summed 
strength measured turned out to agree with theoretical predictions.
 Moreover, the ridge-like shape of strength distribution was observed
except that the position of the peak was found at lower missing energies
than predicted by theory.
 At the same time, Eq.~(\ref{eq:Resc}) predicts little or negligible effects
from FSI in the region of the comparison (see Ref.~\cite{BaLap04} and
discussion of Fig.~\ref{fig:parall_exp} below).
This gives confidence that, for the first time, effects of the high momentum
components attributed to SRC could be observed without being overwhelmed
by the distortion of FSI.
 However, a quantitative understanding of the reaction mechanism
is still needed.

In order to make a meaningful comparison between the experiment and
the theoretical predictions for FSI, one needs a proper ansatz
for the undistorted spectral function, $S^h(k,E)$ in Eqs.~(\ref{eq:Resc})
and~(\ref{eq:PiEm}).
At low energies and momenta we employed the correlated part of
the spectral function $S^h_{LDA}(k,E)$ in Ref.~\cite{Benhar} which was 
obtained using local density approximation (LDA), and combined it with a
properly scaled spectral function $S{^h}_{WS}(k,E)$ derived from an adequate Wood--Saxon potential.
%The parameters of the Wood--Saxon potential were adjusted to previous data. 
This includes the bulk of the spectral strength, located in the mean field 
(MF) region up to a momentum of $\approx$250~MeV/c. 
The remaining strength is located along the SRC ridge.
This represents only a fraction of the total number of nucleons in the system
but it is the the part probed by the kinematics under consideration. Thus,
the most important for the present analysis. 
 As stated above, theoretical calculations can reproduce the qualitative
shape of this distribution and the total summed strength but miss the position
peak~\cite{Frick04}. 
As we are mainly interested in the reaction mechanism, it is preferable
to parameterize the SRC part of the input spectral function in terms of
a Lorentzian energy distribution, which was fitted to the experimental results
for ${}^{12}{\rm C}$ in parallel kinematics~\cite{BaRSL05}.
The full spectral function employed in Eqs.~(\ref{eq:PWIA}--\ref{eq:PiEm})
is~\cite{BaRSL05}
\begin{equation}
 S^h(k,E) = 
  \left\{
      \begin{array}{ll}
         S^h_{Lorentzian}(k,E) & \hbox{for $k > $230~MeV/c} \\
                           & \hbox{and $E > $19~MeV,}   \\
         ~ \\
         S^h_{LDA+WS}(k,E) & \hbox{otherwise.}
      \end{array}
  \right.
 \label{eq:Shtotal}
\end{equation}
In the following, we compare the predictions of the above model
to the experimental data in regions far from the correlated peak and
in different kinematics.

\begin{figure}[!t]
%%\vspace{.15in}
  \begin{center}
    \includegraphics[width=\linewidth]
                                  {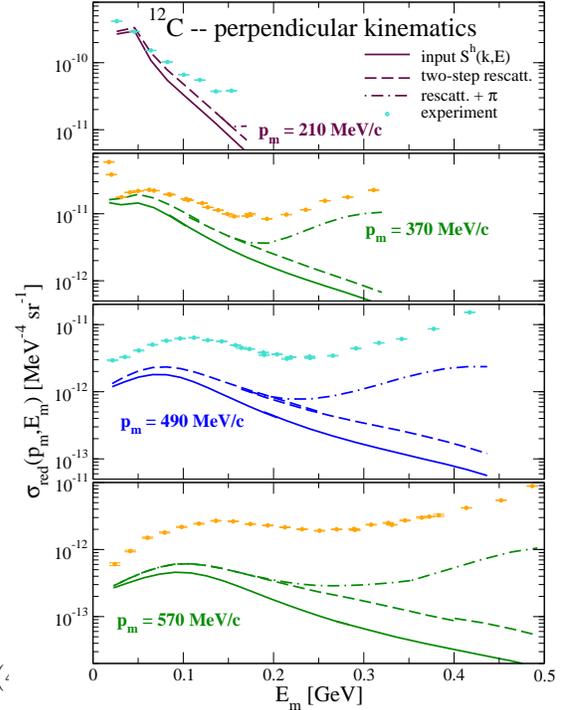}
    \caption{ \label{fig:perp_exp}
       (Color online) 
       Theoretical results for the reduced cross section of
      ${}^{12}{\rm C}$  obtained in
      perpendicular kinematics compared with the experimental
      results of Ref.~\cite{danielaDiss}.
       The full line shows the model spectral function,
      Eq.~(\ref{eq:Shtotal}), employed in the calculations.
       The dashed lines include the additional effects of rescattering,
      Eq.~(\ref{eq:Resc}), while the combined prediction of
      Eq.~(\ref{eq:PWIA}--\ref{eq:PiEm}) is given by the dot-dashed
      lines.
}
\end{center}
\end{figure}

Figure~\ref{fig:parall_exp} compares the results obtained in parallel
kinematics to the experimental reduced cross section defined as
$\sigma_{red}(p_m,E_m)=\sigma_{experiment}/(|p_fE_f|\sigma^{cc}_{ep}{\cal T})$.
When only the PWIA contribution, Eq.~(\ref{eq:PWIA}), is included (full line)
one compares the experiment to the assumed input spectral function $S^h(k,E)$,
showing the quality of the fit around the correlated peak. It is also seen that
the contribution from rescattering processes (dashed line) is negligible
in this region and increases only at very large missing energies~\cite{BaLap04}.
 Thus, validating the choice of taking the input spectral function in this 
region form the experiment.
It is important to observe that the main reason for the small effects
of rescattering  obtained in this calculation is kinematical in origin.
Due to rescattering events, the emitted nucleon loses part of
its initial energy by knocking out a second particle.
 The requirement of small angles between the momenta ${\bf q}$ and
${\bf p}_i$ implies larger energies $T_a$ (i.e. smaller $E_m$)
and larger missing momenta for the intermediate nucleon, with respect
to the detected proton.
This forces sampling the rescattered strength in regions where the
correlated spectral function is smaller than for the direct process.
 For analogous reasons, multiple rescattering effects become even less
important, as seen in Ref.~\cite{VivianPV} for perfectly parallel kinematics.
 Sensible deviation from the Lorentzian shape of the peak are observed
once the pion emission threshold is reached.
 Then, a distorted image of the correlated region is produced at
larger missing energies (due to the extra energy necessary to create
the pion).
 The nice agreement between the total theoretical cross section and the
experiment confirms that in parallel kinematics the contribution of FSI
becomes relevant only at large missing energies and it is dominated by
($e,e'p\pi$) events.
As the  missing momentum increases, the regions dominated by correlated 
nucleons and pion production tend to overlap.
% The small discrepancies at $E_m \approx$200~MeV maybe due to emission of
%pions followed by rescattering, which was not considered here.
%
We note that as long as multiple rescattering effects can be neglected
no sensible quantum interference effect is expected,
since Eqs.~(\ref{eq:PWIA}--\ref{eq:PiEm}) lead to different particles
in their final states.

\begin{figure}[!t]
%%\vspace{.2in}
  \begin{center}
    \includegraphics[width=\linewidth]
                  {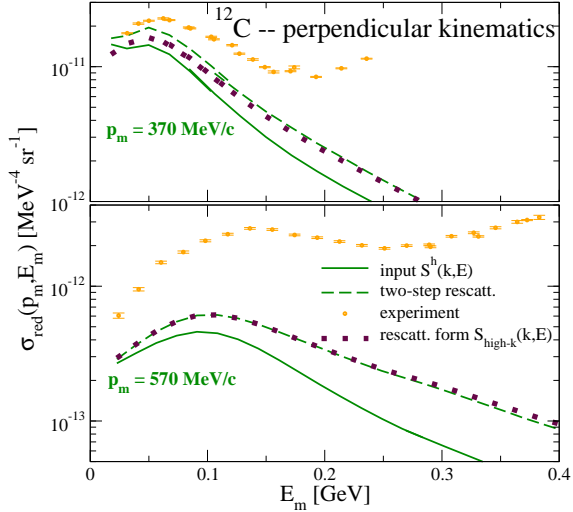}
%\centerline{\psfig{figure=Parall_vs_perp_th.epsi,width=8cm,angle=-90}}
    \caption{ \label{fig:StrRed}
       (Color online) 
      Rescattering contribution to the reduced cross section
      for ${}^{12}{\rm C}$ in  perpendicular kinematics (dashed line).
       The dotted lines show the analogous results obtained by neglecting
      the rescattering from the MF region, $S^h_{LDA+WS}(k,E)$, in
      Eq.~(\ref{eq:Shtotal}).
      % This becomes indistinguishable from the full
      %calculation for $p_m >$~450~MeV/c.
       The full line shows the model spectral function,
      Eq.~(\ref{eq:Shtotal}), employed in the calculations and the experimental
      data are from Ref.~\cite{danielaDiss}
     }
  \end{center}
\end{figure}

The experimental data are found to be sensibly 
larger in perpendicular kinematics than in parallel ones~\cite{danielaDiss}.
 These are compared with
the results of Eqs.~(\ref{eq:Resc}-\ref{eq:PiEm}) in Fig.~\ref{fig:perp_exp}.
In this case, the rescattered strength is large and
affects the reduced spectral function already at the top of the
SRC tail.
Due to the much larger redistribution of spectral strength the valley
between the SRC and meson production regions is also filled in and 
it is no longer possible to distinguish them.
The combined predictions of single rescattering and pion emission are able
to reproduce the shape of the measured cross section but results in sensibly 
lower cross sections.
Fig.~\ref{fig:perp_exp} shows that Eq.~(\ref{eq:Resc}) can account
for the differences between parallel and perpendicular kinematics
for $E_m \approx$~50~MeV and $p_m <$~350MeV/c.
However, the experiment is strongly under predicted over most
of the correlated region.
  For $p_m=$570~MeV/c, the cross section is predicted to be 50\% 
larger than the direct process, whereas the experiment is
almost of one order of magnitude above.
We conclude that two-step rescattering represents only a fraction of the
total FSI present in perpendicular kinematics.
A possible contribution that could bring theory and data closer together
is that of MEC that involve the excitation of a $\Delta$
or higher resonances. These are known to be sensitive to transverse degrees
of freedom.

 We observe that the larger contribution of Eq.~(\ref{eq:Resc}) with
respect to parallel kinematics can also be understood kinematically. 
 In this case a nucleon can lose energy in a rescattering
event but still be detected with a missing momentum larger than its initial
momentum. This results in moving strength from regions where the spectral
function is large to regions where it is small, thereby submerging
the direct signal.
The shift is large enough that measurements in the correlated region can
be affected by events originating from MF orbits~\cite{BaLap04}.
The size of this effect for two-step rescattering is however limited
to the part of the correlated region at low missing momenta.
This is shown in Fig.~\ref{fig:StrRed}, where the dotted lines have been
obtained by setting $S^h_{LDA+WS}(k,E)=0$ in Eq.~(\ref{eq:Shtotal}).
 Single rescattering from the MF strength gives no appreciable contribution
at missing momenta above 500~MeV/c.
For parallel kinematics a similar comparison shows no visible effects
at even lower momenta, indicating that rescattering moves strength {\em only}
within the correlated region itself.
This supports the necessity of including the observed high momentum strength
already in $S^h(k,E)$, that is, in the nuclear wave function.

In conclusion, first predictions of two-step rescattering 
and pion emission in  ($e,e'p$) reacrions at large $p_m$-$E_m$ were
discussed and compared to experimental data~\cite{danielaDiss,danielaPRL}.
 The results are understood in terms of kinematical constraints and
confirm the expectation that, for light nuclei and properly
chosen parallel kinematics, the effects of FSI can be small
for missing energies up to the $\pi$ production threshold.
 In perpendicular kinematics the agreement between data and theory
is less favorable. This suggest that additional ingredients
of transverse character, such as MEC, should be investigated in future
calculations.

%%%%%%%%%%%%%%%%%%%%%%%%%%%%%%%%%%%%%%%%%%%%%%%%%%%%%%%%%%%%%%%%%%%%%%%%%%
%%\acnkowledgments
\vskip .5cm
Several illuminating discussions with D.~Rohe, I.~Sick and L.~Lapik\'{a}s are
acknowledged.
This work was supported by the Natural Sciences and Engineering Research
Council of Canada (NSERC).
%%%%%%%%%%%%%%%%%%%%%%%%%%%%%%%%%%%%%%%%%%%%%%%%%%%%%%%%%%%%%%%%%

\end{document}